\documentclass[aps,showpacs,floats,superscriptaddress]{revtex4-2}

\usepackage{graphicx}
\usepackage{epstopdf}
\usepackage{amssymb}
\usepackage{amsmath,bm}
\usepackage{psfrag}
\usepackage{epsfig}
\usepackage{float}
\usepackage{bm}
\usepackage{textcomp}
\usepackage{hyperref}

\begin{document}
\title{
Model-based study of a nanowire heating and dynamic axisymmetric necking  
by surface electromigration
}

\author{Mikhail Khenner\footnote{Corresponding
author. E-mail: mikhail.khenner@wku.edu.}}
\affiliation{Department of Mathematics, Western Kentucky University, Bowling Green, KY 42101, USA}
\affiliation{Applied Physics Institute, Western Kentucky University, Bowling Green, KY 42101, USA}

\begin{abstract}
\noindent
Axisymmetric solid-state necking 
of a single-crystal metallic nanowire in a thermal contact with a substrate and subjected to a surface 
electromigration current is accompanied by a 
local current crowding and a sharp rise of a resistivity of a wire material in a thinning neck. 
This results in a temperature spike at the neck, which feedback affects the necking via thermomigration and the 
temperature-dependent surface diffusivity of the adatoms. A model that incorporates these effects and couples the nonlinear dynamics 
of a wire temperature and a wire radius for a necking wire is presented. 
Conditions on the physical parameters are derived that ensure a straight wire is in the solid state prior to 
an onset of a morphological instability that ultimately breaks a wire via a pinch-off.
The impacts of a wire radius and a wire length on the temperature spike at the break junction are studied.

\end{abstract}

\date{\today}
\maketitle


\section{Introduction}
\label{Intro}

Electromigration \cite{Huntington,HoKwok,REW} is the major factor adversely affecting the reliability of a microelectronic and nanoelectronic 
interconnects and devices. On the other hand, electromigration also is constructively exploited for creation of nanogaps in a solid thin-film 
conductors, in particular, in a cylindrical nanowires - via a controlled necking and a breakup by pinch-off. 
Placing an organic molecule into a nanogap, its electrical conductance and electronic properties may be measured; however, a 
reliable large-scale fabrication of quality nanogaps using electromigration is still facing challenges \cite{H,KAAC}. 
Controlled breakage of a nanoscale conductive filaments is important also for the emerging field of neuromorphic computing
\cite{WPVSP}.


For fabrication of a high quality electromigration nanogap (a break junction) a wire must be kept away from melting at all times. 
A nanogap is generally far too large if a wire melts at a neck. For a wire to stay solid the current density
through a neck must be reduced proportionally to the rate of a neck thinning. Such control has been achieved either by active feedback 
\cite{SSJPTBJ,EF,HWHS,YS}, 
or by reducing the serial resistance via monitoring the voltage drop over a neck \cite{WSHCMS,TBKR}. Establishing a control is 
technically very difficult, since for a neck thickness smaller than the mean free path of electrons 
($\sim 10-20$ nm or less in the presence of impurities, grain boundaries, etc.), a high temperature up to 
thousands of Kelvins could be triggered by a small voltage rise of the order of 0.1V \cite{HCKMSK}.   

A basic dynamic model of a nanowire necking and breakup on a substrate was introduced by the author in 
Refs. \cite{MyWire1,MyWire2,MyWire3}. 
A single-crystal cylindrical nanowire is considered. 
The basic model assumes that the electric field in the neck is independent of the neck thickness 
(the absence of current crowding) 
and that all physical parameters are constants, i.e. they do not depend on 
the temperature and the neck thickness. Consequently, a wire heating is negligible due to transfer of heat into the substrate 
and the modeling is isothermal. It is based on a partial differential equation
(PDE) that describes the evolution of a wire radius by the surface electromigration-enhanced surface diffusion \cite{NM2,NM1,C,BBW,McCVoorhees,WMVD,SK,GM,MB,WTLZNSMH}. 
The analyses and computations of this PDE resulted in the conditions for the onset of necking and determined the rate of the neck 
thinning and a wire breakup time as a function of major physical parameters.    

In this paper the basic model \cite{MyWire1,MyWire2,MyWire3} is extended by recognizing that the decrease of 
a neck thickness results in the resistivity increase  \cite{JBT,Lacy}, which leads to intense local production of  
Joule heat and thus to the fast rise of the neck temperature, as seen in the experiments \cite{SSJPTBJ,EF,HWHS,YS,HCKMSK,KSL,WHN}
and supported by the steady-state models for the temperature of a thin neck of a given static geometry \cite{KSL,HJFSS,WC,WHN}.
To systematically describe this major effect in the continuum dynamic modeling framework, in this paper the evolution PDE for the wire radius 
is coupled to the initial-boundary value PDE problem for the wire temperature. Simultaneously, the current crowding is included in the statements of both PDEs 
by the appropriate scaling of the electric field parameter with a wire radius \cite{SK,B1,B2}, and the temperature dependence of 
the resistivity is acccounted for.


Growth of metallic 
single-crystal nanowires was achieved on various substrates \cite{SHCHDR,WSRHH,KTEBCKN,YDMGCLS,WZPWMJX}.
Absent electromigration and/or annealing, these nanowires may be grown up to a length of 100 $\mu$m without breaking up. 

For a single-crystal cylindrical nanowire a surface, rather than a bulk, 
electromigration is the major mass transport mechanism that results in a breakup. There is no grain boundary grooving, 
void formation, void coalescence, 
and stress build-up in a single-crystal wire, i.e. absent are many important features that contribute to electromigration 
breakup of a typical mass-produced polycrystalline nanowires. 

Still, the case of a single-crystal wire is important for understanding
a more complicated scenarios for polycrystalline wires, since the surface electromigration and the surface curvature 
(i.e., the surface diffusion) contribute very significantly to the total mass transport at nanoscale. 
The importance of single-crystal nanowires for electromigration studies has been long recognized; for instance, 
Stahlmecke \emph{et al.} \cite{SHCHDR}
point out that ``... the basic electromigration processes must be distinguished
from structural and morphological effects in the
wires. In contrast (to polycrystalline wires), single crystalline wires would provide an
opportunity to study electromigration without structural defects". Hoffmann-Vogel in the excellent review \cite{H} writes:
``In addition, it is known that during the
last stages of the electromigration process, a single grain separated
by grain boundaries is left (Strachan et al., 2008).
Studies of electromigration of single crystals allow us to
focus on the processes that occur in this final stage."

Besides, all model calculations and computations of a neck temperature to-date are for a steady-state temperature of a static 
two-dimensional or three-dimensional nanoconstrictions without defects such as voids and grain boundaries, thus effectively 
they are for single-crystal wires \cite{KSL,WHN,HJFSS,WC}. 
The extended model in this paper contributes the dynamical aspect to these calculations.

In the extended model, termed Model A, not only a progression of a wire necking results in a local temperature spike, 
but owing to thermomigration 
\cite{RC,KJ,XNSYLOLMS,DSGM,LBCMC,CCMBDCPL} and a temperature-dependent surface diffusivity the temperature spike 
feedback affects the necking dynamics. Analysis and computation of this self-consistent coupled dynamical model for a wire 
temperature and a wire radius provides the understanding of the role of the major physical parameters, such as a wire radius 
and length, the applied voltage, etc. in the breakup process and their effect on the temperature of a break junction. 

To complement Model A we also developed a much simpler Model B. In that model the axisymmetric necking is forced by a triangle-shaped surface spike 
with an amplitude which is the input parameter. As a consequence the necking impacts the temperature, but there is no feedback from the
temperature on the necking process. In Sec. \ref{Results} we briefly compare two models.

\emph{
All physical parameters are listed in Table \ref{T1} for reference. These parameters are for gold, but the model 
can use a parameter set that matches any metal with fcc crystal structure and high melting point. 
Since values of $\ell_w$, $R_0$, $V$, and $d_s$ are given in the intervals,  
the actual values are stated in the figure captions. 
}

\section{Model A: Coupled dynamics of a wire temperature and the axisymmetric wire necking}
\label{FullModel}

\subsection{Evolution equations for a wire temperature and radius}

\begin{figure}[H]
\vspace{-0.2cm}
\centering
\includegraphics[width=4.0in]{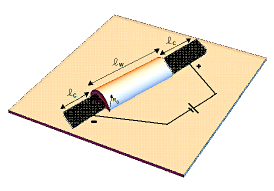}
\vspace{-0.15cm}
\caption{Sketch of a single-crystal metal nanowire and electrical contacts on a substrate. The axial electric current through a 
wire triggers surface electromigration.}
\label{Fig1}
\end{figure}

We consider the temperature evolution via  
Fourier heat conduction in a wire that has the initial semi-circular shape of radius $R_0$ 
and a length $\ell_w$, and in a pair of contacts that are attached to two ends of a wire: 
\begin{eqnarray}
c_w\chi_w \frac{\partial T_w}{\partial t}&=&k_w \frac{\partial^2 T_w}{\partial y^2}-\frac{k_s}{R_0 d_s}T_w+J(t,y)^2 \rho_w(t,y),\quad 0\leq y\leq \ell_w \label{Teqs_dim1}\\
c_c\chi_c \frac{\partial T_c}{\partial t}&=&k_c\frac{\partial^2 T_c}{\partial y^2}-\frac{k_s}{d_c d_s}T_c,\quad -\ell_c\leq y\leq 0\; \cup\; \ell_w\leq y\leq \ell_w+\ell_c. \label{Teqs_dim2}
\end{eqnarray}
Fig. \ref{Fig1} schematically shows a wire-contacts system on electrically insulated substrate. (In reality, the contacts are a massive wide slabs that
dwarf a nanowire \cite{SHCHDR,H,KAAC,YS}.)
$y$ is the axial variable. The subscripts $w$, $s$, and $c$ 
refer to a wire, a substrate, and a contact, respectively;  
the superscripts $(w)$ and $(s)$ refer to a wire and a 
substrate, respectively. The terms $-(k_s/R_0 d_s)T_w$ and $-(k_s/d_c d_s)T_c$ describe the heat losses through the substrate; 
the term $J(t,y)^2 \rho_w(t,y)$ describes the Joule heat production in a wire, where $J(t,y)$ and $\rho_w(t,y)$ are the electric current density and the wire resistivity, 
respectively. Heat generated in the contacts is neglected due to the significantly reduced current
density.
Equations (\ref{Teqs_dim1}) and (\ref{Teqs_dim2}) are motivated by the simpler model discussed in Refs. \cite{Durkan,HJFSS}; that model is a steady-state (time-independent) 
one, where $J$ and $\rho_w$ are a constants. In that model and in our model most of the heat transfer occurs at a nanowire-substrate and a 
contact-substrate interfaces, therefore a substrate acts as a heat sink; the heat losses via a thermal radiation into the ambient 
are negligible compared to the heat losses into a substrate \cite{YDMGCLS}.

The assumption that a wire is semi-circular, i.e. a wire-substrate contact angle is $90^\circ$ is needed to ensure that 
the magnitudes of an axisymmetric morphological instability modes greatly exceed those of a non-axisymmetric ones and therefore 
consideration of the latter may be omitted. Rigorous mathematical analysis and computation is very difficult for contact 
angles other than $90^\circ$ or $180^\circ$ \cite{McCVoorhees,MyWire1} due to the need for consideration of a non-axisymmetric modes, 
however, deviations of the model predictions that are of interest for experiment practice are expected 
to be very minor. If $180^\circ$ contact angle is assumed instead of $90^\circ$, i.e. a full circular wire is considered, then by $R_0$ we understand the diameter of a wire.
(Although the mathematical formulation of the model and its analysis and computation formally hold for such wire, 
we caution that in this    
situation an efficient thermal contact with a substrate is hardly possible, since in this case a wire touches the substrate
along a mathematical line, i.e. a line that has zero thickness.)   

Temporarily suppressing the dependence of $J$ and $\rho_w$ on $t$ and $y$ to simplify writing, 
the term $J^2 \rho_w$ in Eq. (\ref{Teqs_dim1}) is first stated as:
\begin{equation}
J^2\rho_w=\left(\frac{E}{\rho_w}\right)^2\rho_w=\frac{E^2}{\rho_0^{(w)}+\beta_w T_w},
\label{HeatTerm}
\end{equation}
where $E$ is the electric field and $\rho_w=\rho_0^{(w)}+\beta_w T_w$ is the temperature-dependent resistivity 
(with $\beta_w>0$ and $\rho_0^{(w)}\gg \beta_w$). Next, Eq. (\ref{HeatTerm}) is expanded in powers of the 
small parameter $\beta_w/\rho_0^{(w)}$ and linearized, yielding  
\begin{equation}
J^2\rho_w=\frac{E^2}{\rho_0^{(w)}}\left(1-Q_1 T_w\right).
\label{HeatTerm1}
\end{equation}
Here $\quad Q_1=\frac{\beta_w}{\rho_0^{(w)}}$, which we approximate with high precision as 
$\quad Q_1\approx \frac{\beta_w}{\rho_b^{(w)}}$, where $\rho_b^{(w)}$ is the bulk resistivity of a wire material. 

Table \ref{T1} contains twenty five physical parameters that characterize the physical system, which is a large number. 
(This count includes $S$ and $\psi$, which are shown only in the second column of the Table.)
Thus we
introduce the scales for the space variables and the time. This reduces the number of parameters to fourteen, streamlines the formulation, and simplifies 
the analysis and computation. The major results though will be presented in terms of the physical parameters. The following scales are chosen:
\begin{equation}
[\mbox{radius},y]=R_0,\quad [t]=\tau = \frac{R_0^4 k_B T_{op}}{\mathcal{D}\mbox{e}^{-\mathcal{E}/k_B T_{op}}\gamma\Omega^2\nu},
\label{scales}
\end{equation}
where $\tau$ is the standard surface diffusion time scale.
Since a gold wire attached to a copper contacts on SiO$_2$ substrate is assumed as the example system, and a wire temperature 
range is from a room temperature to a melting temperature, in the Arrhenius adatom surface diffusivity entering 
$\tau$ the mean operating temperature 
$T_{op}=T_m/2=532^\circ$C (805K) is chosen. Here $T_m=1064^\circ$C (1337K) is the bulk melting temperature of gold.

The primary contribution $\rho_0^{(w)}$ to the wire resistivity in Eq. (\ref{HeatTerm1}) is introduced next. 
This is modeled after Ref. \cite{Lacy} as follows:
\begin{equation}
\rho_0^{(w)}(r(t,y))=\frac{\rho_b^{(w)}}{g\; r(t,y)(1-\ln{[g\; r(t,y)]})}, \quad g=\frac{R_0}{2s},\quad 0<r(t,y)\leq 1. 
\label{resist}
\end{equation}
Here $s$ is the electron mean free path, $g$ is the dimensionless parameter, and the wire radius $r(t,y)$ is also dimensionless. 
(The unit interval for $r$ corresponds to a physical radius between zero and $R_0$.)
$R_0\ge s$ is assumed, since for smaller radii the classical Fourier heat conduction theory is inadequate. This means that $g\ge 1/2$.
Also it is clear from Eq. (\ref{resist}) that for positivity of 
$\rho_0^{(w)}$ it is necessary that $g\;r(t,y)<e$, and together with $0<r(t,y)\leq 1$ this gives $g<e$. Therefore finally, 
$1/2\le g< e$. Adopting $s=25$ nm \cite{Lacy}, we get $25\;\mbox{nm}=s\le R_0\le 2se\approx 136\;\mbox{nm}$.  
This is the range of values for the wire radius that is considered in the paper.

\begin{figure}[H]
\vspace{-0.2cm}
\centering
\includegraphics[width=3.0in]{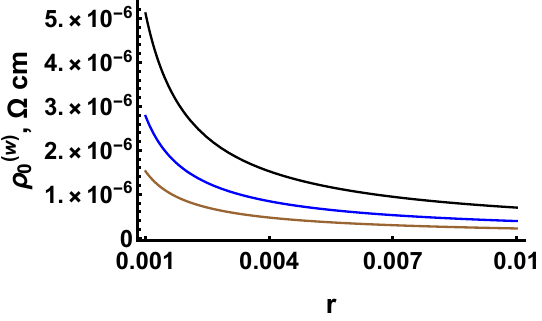}
\vspace{-0.15cm}
\caption{Resistivity, Eq. (\ref{resist}), at $g=1/2, 1, 2$ (black, blue, and brown curves, respectively). }
\label{LacyModelPlot}
\end{figure}

Fig. \ref{LacyModelPlot} shows that at small values of $r$, 
$\rho_0^{(w)}$ increases as $g$ decreases; the latter can be thought of as decreasing $R_0$ at fixed $s$. Notice that at the smallest value $r=0.001$ chosen for plotting, 
which corresponds to a neck thickness 0.025-0.14 nm 
(exact value depends on a chosen value of $R_0$ in the interval stated above), the 
neck resistivity 
is two orders of magnitude larger than the bulk value $2.2\times 10^{-8}$ $\Omega$ cm. 

\emph{Remark 1.} Fully dimensional and $t,y$-independent form of Eq. (\ref{resist}) was suggested by Lacy \cite{Lacy} on the basis of a two-dimensional geometric model, whereby the conduction 
electron is confined to a conductor in the form of an infinite stripe of certain width $R$. The electron travels a distance
of $s$ 
unless it is scattered by the conductor's straight boundaries. Lacy's model 
does not account for the many factors that typically affect the resistivity, such as scattering from grain boundaries, 
scattering from uneven or rough surfaces, and scattering due to impurities. He notes that ``these effects are dependent 
on the procedures and conditions used to fabricate the thin films, and thus, it
is very difficult to quantify each of these effects without
measurement". However, even his base model qualitatively matches the data from the experiment very well \cite{DSMPEBVTA,SVMS}.  

It is seen in Eq. (\ref{resist}) that the dependence of $\rho_0^{(w)}$ on $t$ and $y$ and consequently the proclaimed dependencies of 
$\rho_w$ on $t$ and $y$ are due to a wire radius being a function of these variable during the dynamic necking process. 

The electric field $E$ in Eq. (\ref{HeatTerm1}) is written as:
\begin{equation}
E(r(t,y))=\frac{V/\ell_w}{r(t,y)},
\label{Ecrowd}
\end{equation}
where $V$ is a voltage applied to the contacts and the current crowding due to thinning of a wire is accounted for by dividing the numerator
by the dimensionless wire radius \cite{SK,B1,B2}. Since $\rho_w$ and $E$ are the functions of $t$ and $y$, it follows that the current density 
$J=E/\rho_w$ also is a function of these variables. 

After substitution of equations (\ref{resist}) and (\ref{Ecrowd}) in Eq. (\ref{HeatTerm1}) we substitute the latter equation in 
Eq. (\ref{Teqs_dim1}). Next, we complete adimensionalization of equations (\ref{Teqs_dim1}) and (\ref{Teqs_dim2}) and, 
assuming a steady-state heat conduction in the contacts, 
the final form of the thermal part of the model results:
\begin{eqnarray}
\frac{\partial T_w}{\partial t}&=&D_T^{(w)}\frac{\partial^2 T_w}{\partial y^2}-h_w^{(s)}T_w+q_0(r(t,y))\left(1-Q_1 T_w\right), \quad 0\leq y\leq L_w \label{Teq1_adim}\\
\frac{d^2 T_c^{(\ell)}}{dy^2}-h_c^{(s)}T_c^{(\ell)}&=&0,\quad -L_c\leq y\leq 0 \label{Teq2_adim} \\
\frac{d^2 T_c^{(r)}}{dy^2}-h_c^{(s)}T_c^{(r)}&=&0,\quad L_w\leq y\leq L_w+L_c. \label{Teq3_adim}
\end{eqnarray}
where
\begin{equation}
q_0(r(t,y))=Q_0\frac{1-\ln{[g\; r(t,y)]}}{r(t,y)}. \label{q0}
\end{equation}
Here $Q_0=V^2 \tau g/c_w\chi_w\rho_b^{(w)} \ell_w^2$, 
$D_T^{(w)}=k_w \tau/c_w \chi_w R_0^2$, $h_w^{(s)}=k_s\tau/c_w \chi_w R_0 d_s$, $h_c^{(s)}=k_s R_0^2/k_c d_c d_s$, $L_w=\ell_w/R_0$, 
and $L_c=\ell_c/R_0$. The superscripts $(l)$ and $(r)$ refer to the left and right contacts, respectively.


The following  dimensionless equation describes a non-isothermal evolution of a wire radius during localized 
axisymmetric necking. 
The latter, once initiated due to the onset of a morphological instability,  ultimately results in a wire breakup via a pinch-off.

\begin{equation}
r_t=\frac{1}{r}\frac{\partial}{\partial y} \left[
\mbox{e}^{\frac{\mathcal{\bar E}\left(T_w-T_{op}\right)}{T_w}}
M(r_y)\left\{\frac{r}{\sqrt{1+r_y^2}}\left(K_y+\frac{A}{r^2}+\frac{H\; r\; \partial T_w/\partial y}{T_w\sqrt{1+r_y^2}}\right) \right\}\right], \quad 0\leq y\leq L_w. 
\label{govPDE1}
\end{equation}
Here the subscripts $t$ and $y$ stand for partial differentiation, the exponential factor with 
$\mathcal{\bar E}=\mathcal{E}/k_B T_{op}$ stems from the Arrhenius surface diffusivity,
\begin{equation} 
K=\frac{1}{r\sqrt{1+r_y^2}}-\frac{r_{yy}}{\left(1+r_y^2\right)^{3/2}} \label{curv}
\end{equation}
is the mean curvature of a wire surface, $H=H^* R_0/\gamma \Omega$, and
\begin{equation}
M(r_y)=\frac{1+S\cos^2{\left(\arctan{r_y}+\psi\right)}}{1+S}
\label{mobility}
\end{equation}
is the anisotropic diffusional mobility of the adatoms \cite{SK}. Here $S>0$ is the anisotropy strength and $0\le \psi \le \pi/2$ the misorientation angle 
for a wire oriented along the [110] crystallographic direction. 
Also $A=q V R_0^2/(\Omega \gamma \ell_w)$ is the electromigration parameter, where $q>0$ is the effective charge of ionized adatoms.
The current crowding at the neck is accounted for by the additional $1/r$ factor in $A/r^2$; this mirrors Eq. (\ref{Ecrowd}) \cite{SK,B1,B2}. 
The last term (proportional to $H$)  accounts for 
the thermomigration \cite{DSGM}, which was reported as the contributing factor in electromigration breakup of gold nanowires \cite{YS}.
Motivated by a prior high-quality work \cite{BBW,McCVoorhees,WMVD,SK,GM,MB}, the isothermal model of axisymmetric necking 
without a current crowding and a thermomigration effects 
($\mathcal{\bar E}=H=0$, $A/r^2 \rightarrow A/r$) was introduced by the author in Ref. \cite{MyWire1} and was further studied in 
Refs. \cite{MyWire2,MyWire3}.  

Equations (\ref{Teq1_adim}) and (\ref{govPDE1}) account for all major temperature dependencies and strongly two-way couple the evolution of a 
wire morphology, i.e. the wire radius $r(t,y)$, to the wire temperature $T_w(t,y)$. Since $Q_0>0$ and thus non-zero 
(otherwise the wire heating is impossible) the wire morphology evolution always affects the wire temperature evolution. Setting 
$\mathcal{\bar E}=H=0$ eliminates the feedback effect of the temperature on the morphology evolution via decoupling 
Eq. (\ref{govPDE1}) from (\ref{Teq1_adim}). It must be understood that setting $\mathcal{\bar E}=0$ does not eliminate the surface diffusion 
in this model; this only eliminates the temperature dependence from the surface diffusivity. It is very common in the models 
of this nature to not consider the temperature dependence of the surface diffusivity, accounting for that diffusivity via a constant 
$\mathcal{D}$ (this enters the time scale $\tau$), whose value is chosen according to the mean operating temperature. One glance at Eq. (\ref{govPDE1}) makes it very clear that at $\mathcal{\bar E}>0$ 
(non-zero) the complexity of the equation drastically increases due to differentiation of the triple product and the Chain Rule 
application to the exponent containing $T_w(t,y)$; this makes the computation 
highly costly and especially volatile at the pinch-off stage. Thus the results that we present in Sec. \ref{Results} are primarily 
for the case $\mathcal{\bar E}=0$.

\subsection{Boundary conditions}

Eq. (\ref{govPDE1}) for $r(t,y)$ is fourth-order in $y$. Therefore the boundary conditions at the wire-contact junctions are the symmetry, $r_y=0$, 
and a vanishing total surface diffusion and surface electromigration mass flux. The latter total flux is given by the expression in the square bracket of Eq. (\ref{govPDE1}), where 
the exponent and $M\left(r_y\right)$ are nonzero. Therefore setting to zero the expression in the curly bracket and substituting $r_y=0$ yields the boundary conditions:
 
\begin{eqnarray}
r_y(0)&=&r_y\left(L_w\right)=0,  \label{rsystem1} \\
r_{yyy}(0)&=&\frac{A}{r(0)^2}+\frac{H\; r(0)\; \frac{\partial T_w}{\partial y}(0)}{T_w(0)},  \label{rsystem2} \\
r_{yyy}\left(L_w\right)&=&\frac{A}{r\left(L_w\right)^2}+\frac{H\; r\left(L_w\right)\; \frac{\partial T_w}{\partial y}\left(L_w\right)}{T_w\left(L_w\right)}.  \label{rsystem3}
\end{eqnarray}

The boundary conditions for Equations (\ref{Teq1_adim})-(\ref{Teq3_adim}) reflect the continuity of the 
temperature and the heat flux at a wire-contact junction: 

\begin{eqnarray}
T_w(0)&=&T_c^{(\ell)}(0),  \label{Tsystem1} \\
T_w\left(L_w\right)&=&T_c^{(r)}\left(L_w\right),  \label{Tsystem2} \\
D_T^{(w)} \frac{dT_w}{dy}(0)&=&D_T^{(c)}\frac{dT_c^{(\ell)}}{dy}(0),  \label{Tsystem3} \\
D_T^{(w)} \frac{dT_w}{dy}\left(L_w\right)&=&D_T^{(c)}\frac{dT_c^{(r)}}{dy}\left(L_w\right),  \label{Tsystem4} \\
T_c^{(\ell)}\left(-L_c\right)=T_c^{(r)}\left(L_w+L_c\right)&=&T_a. \label{Tsystem5}
\end{eqnarray}
Here $D_T^{(c)}=k_c \tau/c_c \chi_c R_0^2$ and $T_a$ may be chosen equal to a room (substrate) temperature. 

\emph{Remark 2.} 
If $q_0(r(t,y))$ 
in Eq. (\ref{Teq1_adim}) were a constant,
then the linear problem (\ref{Teq1_adim})-(\ref{Teq3_adim}), (\ref{Tsystem1})-(\ref{Tsystem5})
would admit an analytical solution. However, $r(t,y)$ in $q_0$ is a complicated function of both variables that is \emph{a priory} unknown; 
it is a solution of a highly nonlinear evolution Eq. (\ref{govPDE1}) for the wire radius. Moreover, the coefficients in Eq. (\ref{govPDE1})
are the functions of $T_w$, i.e. equations (\ref{Teq1_adim}) and (\ref{govPDE1}) are two-way coupled.
Thus the solution of Eq. (\ref{Teq1_adim}) may be only numerical.

\section{Model B: Wire heating in the arbitrary proximity to a pinch-off }
\label{ModelSimple}

In this section we aim to determine a steady-state temperature distribution in a wire 
with the help of a pure geometric model of axisymmetric necking. The latter model would have an adjustable input parameter that 
controls the neck thickness. 

To begin, let the the coefficient of $1-Q_1 T_w$ in Eq. (\ref{Teq1_adim}) be 

\begin{equation}
q_0(r(y))=Q_0\frac{1-\ln{[g\; r(y)]}}{r(y)}. \label{barq0}
\end{equation}
Here the time dependence of $r$ is suppressed. Then a steady-state Eq. (\ref{Teq1_adim}) reads:
\begin{equation}
D_T^{(w)}\frac{\partial^2 T_w}{\partial y^2}-h_w^{(s)}T_w+q_0(r(y))\left(1-Q_1 T_w\right)=0, \quad 0\leq y\leq L_w. \label{Teq1_adim_ss}
\end{equation}
Using the steady-state approximation of Eq. (\ref{Teq1_adim}) tacitly assumes that the temperature responds fast to a changes of a wire radius.

We model the neck by the ``spiked" form of $r(y)$, choosing the latter as the following piecewise linear function that conserves a wire volume.
The graph of $r(y)$ is symmetric about the line $y=L_w/2$:
\begin{equation}
r(y)=\left\{
\begin{array}{ccc}
1+d+\frac{1}{2}\left(\alpha L_w-\sqrt{\alpha L_w\left(4d+\alpha L_w\right)}\right), & 0\leq y \leq \frac{L_w}{2\sqrt{1+4d/\alpha L_w}}  \\
1+d+\frac{1}{2}\alpha L_w-\left(\frac{4d}{L_w}+\alpha\right)y, &  \frac{L_w}{2\sqrt{1+4d/\alpha L_w}} \leq y \leq \frac{L_w}{2}\\
1-3d-\frac{1}{2}\alpha L_w+\left(\frac{4d}{L_w}+\alpha\right)y, & \frac{L_w}{2} \leq y \leq L_w \left(1-\frac{1}{2\sqrt{1+4d/\alpha L_w}}\right) \\
1+d+\frac{1}{2}\left(\alpha L_w-\sqrt{\alpha L_w\left(4d+\alpha L_w\right)}\right), & L_w \left(1-\frac{1}{2\sqrt{1+4d/\alpha L_w}}\right) \leq y \leq L_w
\end{array}
\right.
\label{rVpreserve}
\end{equation}
Here the input parameter $0<d<1$ is the spike ``amplitude" and $\alpha>0$ is the parameter that sets the (positive) slope of the right wall of a spike:
\begin{equation}
m=\frac{4d}{L_w}+\alpha.
\label{slopedepth}
\end{equation}

\begin{figure}[H]
\vspace{-0.2cm}
\centering
\includegraphics[width=6.0in]{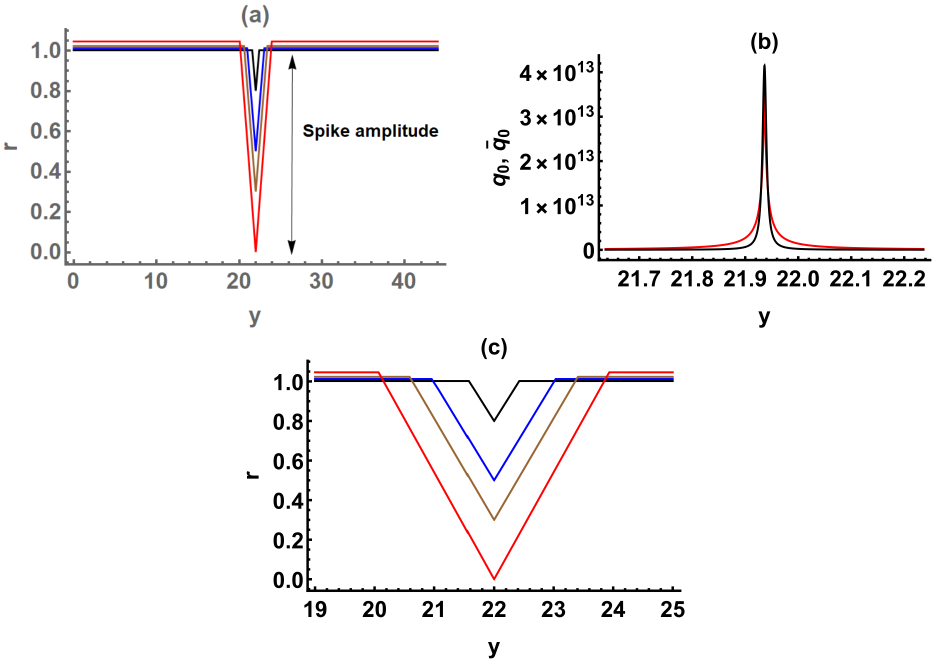}
\vspace{-0.15cm}
\caption{(a) Example spike profiles at various amplitudes. Black, blue, brown, red lines: $d=0.2,\ 0.5,\ 0.7,\ 0.999$, respectively. 
(b) Zoom view of $q_0(r(y))$ (red curve) and $\bar q_0(r(y))$ (black curve) at $d=0.997$. 
(c) Zoom view of the spike profiles in (a). 
$L_w=44$, $\alpha=0.55$, $d_s=300$ nm, $V=0.2$ V, $g=1$.}
\label{Fig2}
\end{figure}

Computation of Eq. (\ref{Teq1_adim_ss}) would incur the smallest numerical error whenever the heat term is continuously differentiable, 
therefore after Eq. (\ref{rVpreserve}) is substituted in Eq. (\ref{barq0}), $q_0(r(y))$ is interpolated by a (smooth) 
Runge function that has the same range: 
\begin{equation}
q_0(r(y)) \approx \bar q_0(r(y))=q_{0,min}+\frac{q_{0,max}-q_{0,min}}{1+a\left(y-L_w/2\right)^2},\quad 
q_{0,min}=\min_{0\leq y\leq L_w} q_0(r(y)), \quad q_{0,max}=\max_{0\leq y\leq L_w} q_0(r(y)).
\label{RungeFit}
\end{equation}
Here $a>0$ is a fit parameter. The values $q_{0,min}$ and $q_{0,max}$ are calculated prior to the smoothing by a Runge function.   

Despite the simplification of Eq. (\ref{Teq1_adim_ss}) via the approximation (\ref{RungeFit}), 
that equation still can't be solved analytically and thus a solution will be computed. 

Fig. \ref{Fig2} shows the spike and the comparison of the graphs of $q_0(r(y))$ and $\bar q_0(r(y))$.

\emph{Remark 3.} Since $Q_1\sim 10^{-6}$ and $T_w$ is capped by $T_m=1337$K (see Sec. \ref{LSA}), $Q_1 T_w\ll 1$ and one may omit 
$Q_1 T_w$ term from Eq. (\ref{Teq1_adim})
without significantly affecting the temperature profiles and the temperature at the spike tip. 
Omitting $Q_1 T_w$ term in Model B 
leads to the analytical solution of Eq. (\ref{Teq1_adim_ss}), however that solution is not in a convenient or insightful form, as 
it is in terms of the exponential integral function due to Eq. (\ref{RungeFit}). 
Omitting $Q_1 T_w$ term in Model A one still needs to compute Eq. (\ref{Teq1_adim}).
Thus there is no benefit in approximating $Q_1 T_w$ by zero and this term is retained. 

The results of Model B will be compared to Model A in Sec. \ref{Results}. To best match the shape of the spike in Model B to the one in Model A,
first, $d$ is fixed to a value reached in the computation of Model A. Next, the parameter $\alpha$ in Eq. (\ref{slopedepth}) is chosen such that the width of 
the spike at a half-minimum equals the width of the spike at a half-minimum in Model A.

The advantage of Model B is that 
the spike amplitude $d$ may be chosen arbitrarily close to one, thus the pinch-off event (i.e., a zero wire radius) 
may be approximated to arbitrary precision.

\section{Base State Wire Temperature via the Linear Stability Analysis of Model A
}
\label{LSA}

In this section a relation between a wire radius $R_0$ and its length $\ell_w$ is established, such that at fixed applied voltage
a wire is a single-crystal solid prior to the onset of a morphological instability.  Then, a fastest growing wavelength of a small destabilizing 
axial perturbation of a wire is determined. In Sec. \ref{Results} it is shown that with the passage of time
this growth develops into a thinning neck which culminates in a wire pinch-off.

\begin{figure}[H]
\vspace{-0.2cm}
\centering
\includegraphics[width=3.0in]{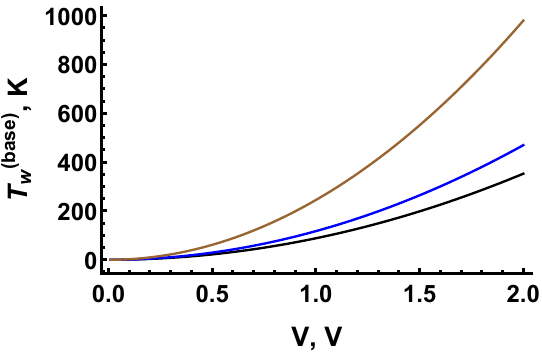}
\vspace{-0.15cm}
\caption{Base temperature of a wire at $\ell_w=7500$ nm (black curve), $\ell_w=6500$ nm (blue curve), 
and $\ell_w=4500$ nm (brown curve). $R_0=50$ nm, $d_s=300$ nm.}
\label{Twbase_vs_V}
\end{figure}

Equations of Model A
admit a base state solution that corresponds to a wire of a constant initial thickness $R_0$ along its entire length 
and a constant temperature:
\begin{equation}
r(t,y)=r^{(base)}=1,\quad T_w(t,y)=T_w^{(base)}
=\frac{\ln{g}-1}{Q_1\left(\ln{g}-1\right)-\delta},\label{Twbase}
\end{equation}
where
\begin{equation}
\delta=\frac{h_w^{(s)}}{Q_0}=\frac{k_s \rho_b^{(w)} \ell_w^2}{g R_0 d_s V^2}. \label{Twbase_aux}
\end{equation}
Note that $T_w^{(base)}$ does not depend on $\mathcal{\bar E}$ and $H$, also recall that $g=R_0/2s$. In Fig. \ref{Twbase_vs_V} 
$T_w^{(base)}$ is plotted vs. the applied voltage for several wire lengths. These curves have the shape very similar to one 
computed for gold nanowires by Aherne \textit{et al.} \cite{ASF}.

\begin{figure}[H]
\vspace{-0.2cm}
\centering
\includegraphics[width=6.0in]{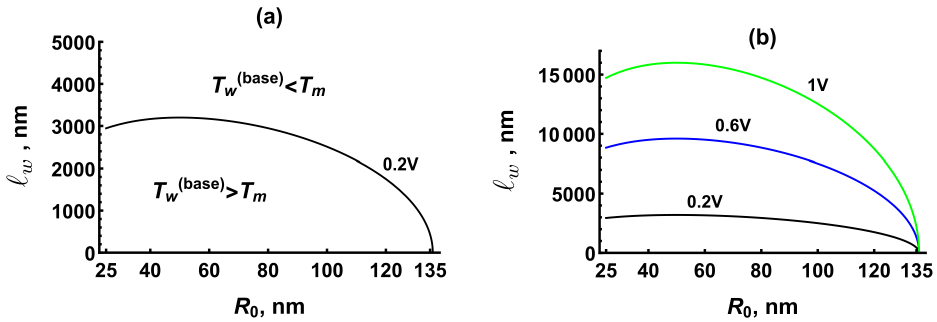}
\vspace{-0.15cm}
\caption{(a) A melt curve corresponding to a fixed applied voltage (marked), that separates the region where the wire base temperature is lower than the melting temperature 
from the region where the wire base temperature is higher than the melting temperature. 
(b) Melt curves at the three values of the applied voltage. $d_s=300$ nm.}
\label{FigTwbase}
\end{figure}

After setting $T_w^{(base)}=T_m$ we find the ``melt" curve 
$\ell_w\left(R_0\right)$, which separates the domains in the $R_0-\ell_w$ plane where $T_w^{(base)}<T_m$ and $T_w^{(base)}>T_m$ 
(Fig. \ref{FigTwbase}(a)). Only the former domain
has the physical meaning in the model, since a wire is assumed solid 
and single-crystalline at all times.
It can be seen that up to $R_0\approx 130$ nm,  $T_w^{(base)}>T_m$ for wire lengths exceeding the classical Rayleigh-Plateau threshold 
$\ell_w=2\pi R_0$ of 
a morphological instability without electromigration and thermal effects (whereby, the mass transport is only by isothermal surface diffusion), before $T_w^{(base)}$ drops below $T_m$ at large $\ell_w$.   
In Sec. \ref{Results} the wire radius $R_0$ and its length $\ell_w$ are varied on the intervals, such that $T_w^{(base)}<T_m$ as in Fig. \ref{FigTwbase}(a) 
and a long-wavelength instability as described next takes place. 

In Fig. \ref{FigTwbase}(b) three melt curves are plotted that correspond to three values of the applied voltage. 
As the voltage increases the melt region grows. It is also seen from Equations (\ref{Twbase}), (\ref{Twbase_aux}) that 
$T_w^{(base)}\rightarrow 1/Q_1\sim 10^6\gg T_m$ as $V\rightarrow \infty$ and other parameters are fixed. 

\begin{figure}[H]
\vspace{-0.1cm}
\centering
\includegraphics[width=3.0in]{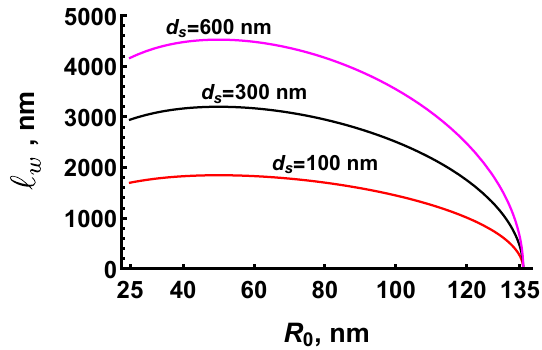}
\vspace{-0.15cm}
\caption{Melt curves at the three values of the substrate thickness $d_s$. The black curve is the same as in Fig. \ref{FigTwbase}.
$V=0.2V$.}
\label{FigTwbase1}
\end{figure}

It was demonstrated that the properties of the thermally conductive substrate are important for control of the 
film temperature and dynamics \cite{HJFSS,ACK}. In our model there is two relevant parameters, the substrate thermal conductivity $k_s$ 
and the substrate thickness $d_s$. They enter the dimensionless parameters $h_w^{(s)}$, $h_c^{(s)}$, and $\delta$ in the combination $k_s/d_s$. 
Thus in fact only the ratio affects the solutions, and not each parameter separately. 
We assume for a moment that $k_s$ is still fixed, i.e. the substrate material is the same (SiO$_2$), but the substrate 
thickness $d_s$ is varied, and in Fig. \ref{FigTwbase1} we show the melt curves at the applied voltage $0.2V$ and the three 
values of $d_s$. 

\emph{Remark 4.}
The equation of a melt curve has the form 
$\ell_w=V \sqrt{\left[a_1+a_2 \ln{R_0}\right]R_0 d_s}$, where $a_1$ and $a_2$ mark numerical values that depend 
on the physical parameters other than $\ell_w,R_0,V$ and $d_s$. This shows that on a melt curve $\ell_w \sim V\sqrt{d_s}$. 

Omitting the boundary conditions, perturbing $r^{(base)}$ and $T_w^{(base)}$ on 
$-\infty < y < \infty$ by a small perturbations $\xi(t,y)$ and $\eta(t,y)$, respectively 
and 
linearizing Eqs. (\ref{Teq1_adim}) and (\ref{govPDE1})  gives the equations for the perturbations in the form

\begin{equation}
\xi_t= F\left(\xi_y,\xi_{yy},\xi_{yyyy},\eta_y,\eta_{yy}\right),\quad \eta_t= G\left(\xi,\eta,\eta_{yy}\right).
\end{equation}


Substituting the normal modes
\begin{equation}
\xi=u\mbox{e}^{\left(\omega_r + i \omega_i\right) t+iky},\quad \eta=v\mbox{e}^{\left(\omega_r + i \omega_i\right) t+iky},
\end{equation}
and separating the real and imaginary parts yields

\begin{equation}
S_1
\begin{bmatrix}
u\\
v
\end{bmatrix}
=
\begin{bmatrix}
0\\
0
\end{bmatrix},\quad
S_2
\begin{bmatrix}
u\\
v
\end{bmatrix}
=
\begin{bmatrix}
0\\
0
\end{bmatrix},
\label{mform}
\end{equation}
where
\begin{equation}
S_1=
\begin{bmatrix}
\omega_r+p_1k^2+p_2k^4 & p_3k^2\\
p_4 & \omega_r+p_5k^2+p_6
\end{bmatrix},\quad
S_2=
\begin{bmatrix}
\omega_i+p_7 k & p_8 k\\
0 & \omega_i
\end{bmatrix}
\end{equation}
and $p_1-p_8$ are a complicated expressions involving the dimensionless parameters. Requiring a unique solution of Eqs. (\ref{mform})  
gives:

\begin{eqnarray}
\mbox{Det}(S_1)&=&\omega_r^2+\left[p_2 k^4+\left(p_1+p_5\right)k^2+p_6\right]\omega_r+
\left(p_1p_6-p_3p_4\right)k^2+\left(p_2p_6+p_1p_5\right)k^4+p_2p_5k^6=0, \label{Det1}\\
\mbox{Det}(S_2)&=&0\quad \rightarrow \quad \omega_i(k)=
-k p_7=-k A \frac{2+S\left(1+\cos{2\psi}\right)}{2(1+S)}\mbox{e}^{\frac{\mathcal{\bar E}\left(T_w^{(base)}-T_{op}\right)}{T_w^{(base)}}}.
\label{Det2}
\end{eqnarray}

Two branches of the real part $\omega_r(k)$ of the perturbations growth rate $\omega(k)$ are determined as the solution of the quadratic 
Eq. (\ref{Det1}), whereas the imaginary part $\omega_i(k)$ seen in Eq. (\ref{Det2}) is unique. From that equation one notes that 
the perturbation drift velocity along the wire axis is $p_7$, which is proportional to the applied voltage parameter 
$A$ and does not depend on a perturbation wavenumber $k$ \cite{SK}. 
At $R_0$ and $\ell_w$ satisfying $T_w^{(base)}<T_m$, one branch of $\omega_r(k)$ 
is negative at $k>0$, and another has a typical form for a long-wavelength instability, that is,
$\omega_r = 0$ at $k=0, k_c$, $\omega_r>0$ on $(0,k_c)$, $\omega_r<0$ on $(k_c,\infty)$ and $\omega_r'=0$ at $k=k_m$, 
where $0<k_m<k_c$. $\lambda_m=2\pi/k_m$ is a wavelength of a fastest growing instability mode.

\section{Remarks on the computation setup and the parameters}
\label{Notes}

We compute equations of Model A using fourth-order finite differences in space on a fixed grid and a variable order 
backward difference method in time. For the accurate and stable computation at a spike tip a very fine grid is required. 
Our computer hardware supports a maximum total of about 2500-3000 grid points, which results in a coupled system of 5000-6000 nonlinear ODEs
after a spatial discretization.

We compute evolution of the spike and the temperature starting from a random, small amplitude initial perturbations about the base state.
Therefore due to a perturbations coarsening and their drift, a wire pinch-off occurs at an unpredictable location along a wire axis.
Each computer run is terminated when the spike amplitude reached 0.997, that is, the wire radius at a pinch-off is 0.003. 
Depending on $R_0$, this translates to the dimensional wire radius at a pinch-off that equals to 0.53-1.5 radius of a gold atom, 
meaning that a ``true" pinch-off event can be declared. The shapes of a spike and a temperature profiles and the temperature at 
a spike tip, $T_{spike}$ are recorded at a pinch-off.  

Dimensionless applied voltage parameter $A$ in the radius evolution Eq. (\ref{govPDE1}) is proportional to the product of the applied voltage, the adatom effective charge, 
and the square of a wire radius, and it is inversely proportional to the square of a wire length.
This parameter needs a tight control due to the following complicating factors:
a perturbation drift with the velocity that is proportional to $A$, 
the need to ``follow" a spike to a pinch-off that has to occur inside a finite computational domain, 
and the need for high accuracy at a spike tip, which calls for that domain to be small, i.e. equal to a small integer multiple of the most 
dangerous wavelength $\lambda_m$. 
Therefore we chose to fix $A=0.1$, which satisfies all stated objectives. Keeping $A$ constant may the thought of as adjusting the 
applied voltage in response to a change of a wire radius or a wire length. (Not keeping $A$ constant, i.e. re-computing 
this parameter in response to a change of a wire radius or a wire length would affect the speed of the necking, the neck location, 
the shape of a wire, and the rate of a temperature rise at the neck, but this would not affect 
the final (i.e., at the time of a pinch-off) magnitude of a temperature spike, 
which is the main object of the computation.)    
Also, values of the thermomigration heat of 
transport $H^*$ are few and vary wildly in the experiment publications, thus instead of choosing a range of values for $H^*$ we opted to 
vary the dimensionless parameter $H$ from zero to a maximum value that is empirically chosen so that the computation stays stable 
and reliable up to a pinch-off. Likewise, instead of using the value of $\mathcal{\bar E}$ in Table \ref{T1}, we choose either   
$\mathcal{\bar E}=0$ or $\mathcal{\bar E}=7\times 10^3$, which is the maximum value that a stable and reliable computation sustains 
(corresponding to $\mathcal{E}=0.606$ eV). 
A reader should also refer to the remarks on $\mathcal{\bar E}$ in Section \ref{FullModel}.   

Excluding $A$, the dimensionless parameters that depend on $R_0$ or $\ell_w$ are re-calculated 
each time these parameters change values, then a
corresponding $\lambda_m$ is calculated and $L_w$ is set to $5\lambda_m$. We also calculate new $\lambda_m$ and set $L_w=5\lambda_m$ 
when $H$ is adjusted, or when $\mathcal{\bar E}$ is chosen either zero or $7\times 10^3$.  

\section{Computational Results}
\label{Results}

\begin{figure}[H]
\vspace{-0.2cm}
\centering
\includegraphics[width=6.0in]{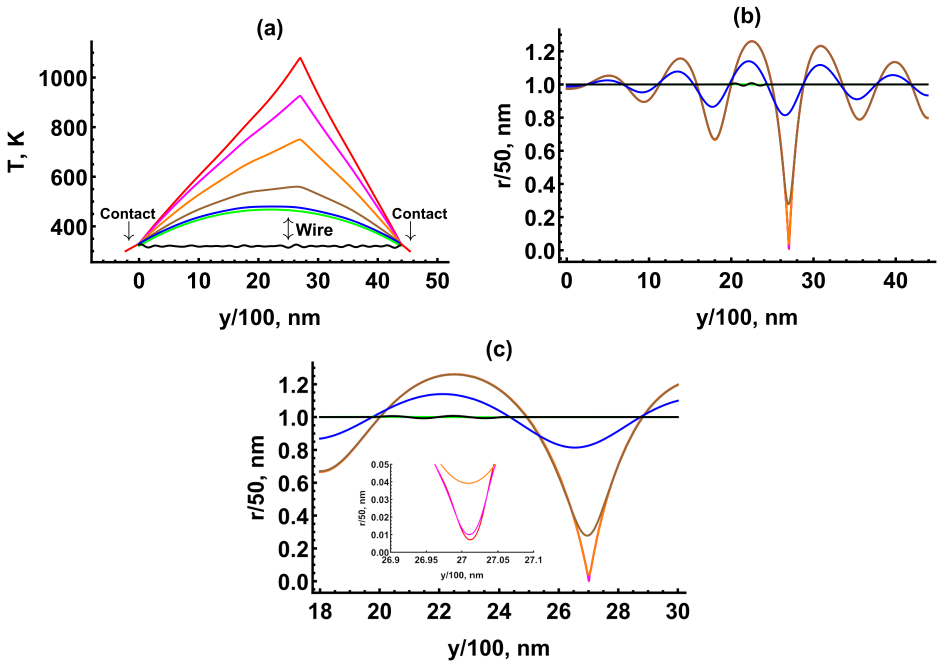}
\vspace{-0.15cm}
\caption{(a) Wire temperature at the 
increasing times.  The times are, bottom to top: $t=0,1,24,26.7,26.72903,26.72903975,26.729039773$.
The bottom-most curve is the initial condition.
(b) Spike profiles at the same times, the curves colors match the colors in (a).
(c) Zoom views of the spike profiles. $t=26.729039773$ is the assumed pinch-off time, i.e. this is the time by which the surface spike amplitude reached 0.997.
$\ell_w=4200$ nm, $R_0=50$ nm, $d_s=300$ nm, $V=0.2V$, $H=\mathcal{\bar E}=0$.}
\label{Fig3}
\end{figure}

For the parameters in Fig. \ref{Fig3} caption, that Figure shows the wire temperature and the surface spike profiles at the increasing times (Model A). 
This is shown only as the example; for other parameters the dynamics of the temperature and the morphology are qualitatively similar. 
The wire heats up slow until the surface spike starts accelerating toward a pinch-off. From this point on the temperature rises fast. 
The temperature spike starts forming around $t=26.7$ (the brown curve), simultaneously with the surface spike.
The asymmetry of the temperature profiles reflects the surface spike asymmetry (see the inset in (c)). The maximum temperature at $t=26.729039773$,
when the surface spike amplitude reaches the pinch-off value 0.997, is 1114K. In this simulation the wire is not quite close to melting at the tip of the surface spike.

The computed wire breakup via the extension of an axisymmetric surface spike to the substrate resembles 
a surface electromigration-triggered propagation of erosion/slitlike voids across 
single-crystal micrometer-size metallic lines \cite{JT,GM,MB}. In a similar fashion, the surface spike not only grows but also moves 
along the wire driven by the electric field. We also note that the computed temperature profiles in Fig. \ref{Fig3}(a) 
(showing sharpening at the later stages of the wire necking) are similar to the calculated ones for the static two-dimensional bow tie 
nanoconstrictions \cite{WC}. Large temperature of around 1000 K at the surface spike tip, i.e. at the most narrow point of a wire, 
was directly measured in the experiment on electromigration breakup of Pt nanowires \cite{WHN}.    

\begin{figure}[H]
\vspace{-0.2cm}
\centering
\includegraphics[width=5.5in]{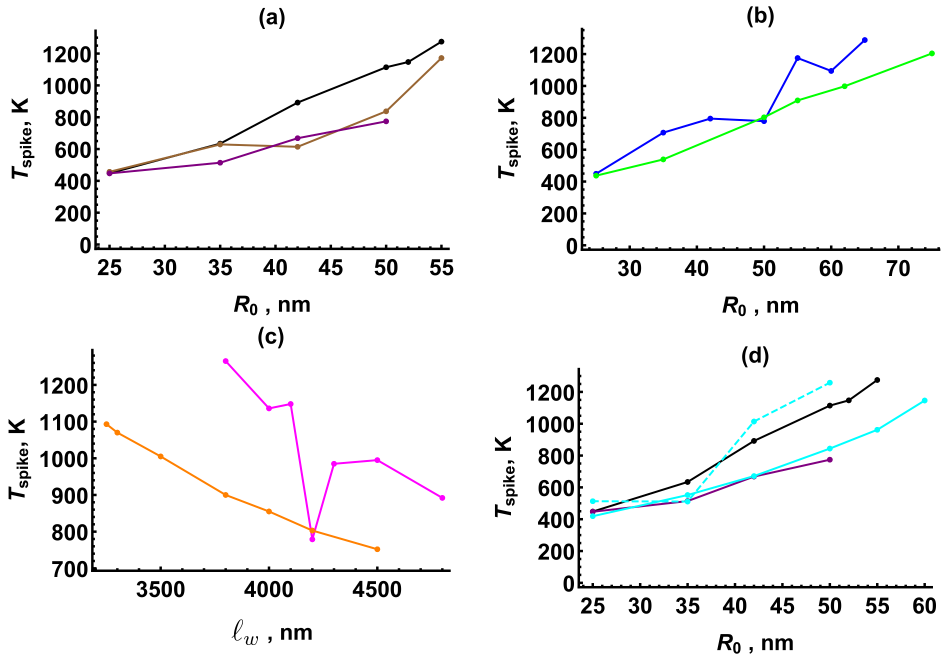}
\vspace{-0.15cm}
\caption{(a,b,d): Temperature at the spike tip at the pinch-off spike amplitude 0.997 vs. $R_0$. $\ell_w=4200$ nm, $d_s=300$ nm, $V=0.2V$. 
(a) Black curve: $H=0$, brown curve: $H=5$, purple curve: $H=50$; $\mathcal{\bar E}=0$. 
(b) Blue curve: $\mathcal{\bar E}=0$, green curve: $\mathcal{\bar E}=7\times 10^3$; $H=1$. 
(d) Black and purple curves are the copies from (a). Solid cyan curve is Model B at the spike amplitude there $d=0.997$, 
which matches the final amplitude of Model A; dashed cyan curve is Model B at the spike amplitude there $d=0.9999$. 
(c): Temperature at the spike tip at the pinch-off spike amplitude 0.997 vs. $\ell_w$. $R_0=50$ nm, $d_s=300$ nm, $V=0.2V$, $H=1$. Magenta curve: 
$\mathcal{\bar E}=0$, orange curve: $\mathcal{\bar E}=7\times 10^3$. }
\label{Fig4}
\end{figure}

In Fig. \ref{Fig4} the temperature at the spike tip at the pinch-off time is plotted vs. $R_0$ and $\ell_w$ (Model A). 

Panel (a) shows that 
$T_{spike}$ rises as $R_0$ increases, and decreases as $H$ increases for $R_0<50$ nm. Notice how at $R_0>50$ nm 
the curves rise more steeply as $H$ increases. For $H=50$ (the purple curve), $T_{spike}>T_m$ at $R_0>50$ nm 
and the curve is not continued beyond $R_0=50$ nm. The maximum temperature difference between three $H$ cases in panel (a) is 250K,
and the maximum temperature difference $T_{spike}(55 \mbox{nm})-T_{spike}(25 \mbox{nm})=830$K is for the the black curve ($H=0$ case).  

In panel (b) $\mathcal{\bar E}=0$ and $\mathcal{\bar E}> 0$ cases are compared. For the former case $T_{spike}$ rises faster as $R_0$ increases, and the 
wire is considered melted at $R_0>65$ nm.

In panel (c) $T_{spike}$ is plotted vs. wire length at fixed $R_0=50$ nm, fixed $H=1$, and again for $\mathcal{\bar E}=0$ and $\mathcal{\bar E}> 0$.
Although the temperature decreases as $\ell_w$ increases in both cases, there is significant differences. In the latter case 
the average temperature is much higher and at $\ell_w<3800$ nm the wire melts. In the former case there is the surprising and 
large temperature drop of around 360K at $\ell_w\approx 4200$ nm and at $\ell_w<3250$ nm the wire melts. One can also notice in panels (b) and (c) that letting 
$\mathcal{\bar E}> 0$ has a smoothing effect on the temperature variations.

Panel (d) shows that Model B significantly underestimates $T_{spike}$. At the same amplitude $d=0.997$ as in Model A the prediction for $T_{spike}$ 
from Model B is comparable to the prediction from Model A only at $H=50$ there. The largest difference between Model A at $H=0$ and Model B
is at $R_0=50$ nm (270 K), and the average difference is 153 K. However, Model B allows to reach the spike amplitudes that are presently not accessible with the 
computation of Model A. 
For instance, adjusting $T_{spike}$ values on the dashed cyan curve (corresponding to the amplitude $d=0.9999$ in Model B, 
i.e. the neck thickness is essentially zero) 
upward by the stated average 
difference value 153 K, predicts that at $\ell=4200$ nm a wire melts at the tip of the spike when $R_0\approx 47$ nm.

\begin{figure}[H]
\vspace{-0.2cm}
\centering
\includegraphics[width=3.0in]{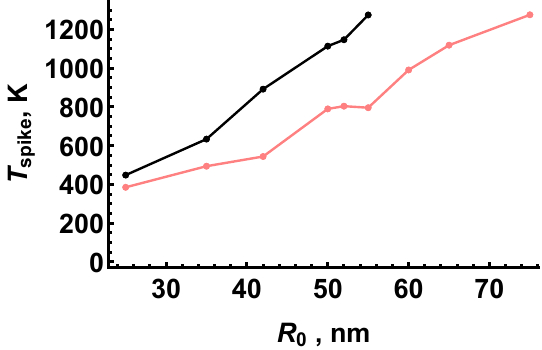}
\vspace{-0.15cm}
\caption{Temperature at the spike tip at the pinch-off spike amplitude 0.997 vs. $R_0$.  
$\ell_w=4200$ nm, $R_0=50$ nm, $V=0.2V$, $H=\mathcal{\bar E}=0$. 
Black curve: $d_s=300$ nm (copy from Fig. \ref{Fig4}(a)). Pink curve: $d_s=100$ nm.
 }
\label{Fig5}
\end{figure}

Fig. \ref{Fig5} shows the comparison of the temperature at the spike tip at the pinch-off time for two values of the substrate 
thickness (Model A). Larger substrate thickness results in faster growth of $T_{spike}$.

\begin{figure}[H]
\vspace{-0.2cm}
\centering
\includegraphics[width=3.0in]{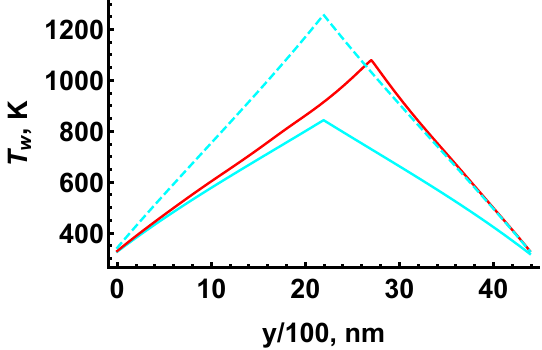}
\vspace{-0.15cm}
\caption{Wire temperature at a pinch-off. Red curve is the copy from Fig. \ref{Fig3}(a) (Model A). Solid cyan curve is Model B at $d=0.997$, dashed 
cyan curve is Model B at $d=0.9999$. $\ell_w=4200$ nm, $R_0=50$ nm, $d_s=300$ nm, $V=0.2V$, $H=\mathcal{\bar E}=0$ 
(the latter two values are irrelevant for Model B). }
\label{Fig6}
\end{figure}

Lastly, Fig. \ref{Fig6} compares the wire temperature profiles in Models A and B. The latter profiles are symmetric about the line $y=\ell_w/2$. 
The former profiles are asymmetric, with a maxima that are shifted with respect to the line $y=\ell_w/2$ due to a lateral surface drift.
It is again obvious that Model B underestimates the temperature.

\section{Conclusions}

Presented model of a non-isothermal, axisymmetric solid-state necking by surface electromigration resulted 
in a relation between a wire length and a wire radius, maintaining which at a fixed applied voltage keeps the temperature of 
an initial semi-circular (undeformed) wire below the melting temperature. 
In particular, it is shown that the base state 
temperature for a fairly short and thin wire is above the melting temperature, i.e. such wire is a liquid filament prior to the 
onset of any morphological instability. In this category fall wires whose length to radius ratio significantly exceeds the 
classical Rayleigh-Plateau ratio $2\pi$ that in a classical linear theory is necessary for the onset of a morphological instability. 
On the other hand, the base state of a thick and short wire is a solid. 
Excluding from consideration the lengths and radii such that the base state of a wire is a liquid, 
for a solid wire the progressive solid-state necking and the associated 
locally sharpening temperature spike were computed until a wire nearly pinches off.  The influences of a wire radius, wire length, substrate thickness, 
and a thermomigration strength on the temperature spike were computed. The increase of a wire radius at fixed wire length results 
in the increase of a neck temperature, in some cases going beyond the melting temperature. The increase of a wire length at fixed wire 
radius results in the decrease of a neck temperature. Thermomigration decreases the neck temperature at a relatively small wire radii,
but sharply increases the neck temperature at a large wire radii. A few results are contrasted to the ones from a reasonable simpler 
model, whereby the necking and the temperature rise are quasistatic and the necking affects the temperature, but the temperature 
does not affect the necking. It is shown that such model significantly 
underestimates the neck temperature, accentuating the need for a coupled multi-physics dynamical models of a wire necking and 
a wire temperature for a future theoretical support of a nanowire electromigration experiments.        

\begin{table}[!ht]
\centering
{\scriptsize 
\begin{tabular}
{|c|c|c|c|}

\hline
				 
			\rule[-2mm]{0mm}{6mm} \textbf{Physical parameter} & \textbf{Value}	 & \textbf{Dimensionless parameter} & \textbf{Value}\\
			\hline
                        \hline
			\rule[-2mm]{0mm}{6mm} Wire radius, $R_0$ & 25-136 nm &  Wire thermal diffusivity, $D_T^{(w)}$ & $1.29\times 10^{10}$ (*)\\
			\hline
			\rule[-2mm]{0mm}{6mm} Wire length, $\ell_w$ & 3000-5000 nm & Wire length, $L_w$ & 84  (*)(see Sec. \ref{Notes})\\
			\hline
			\rule[-2mm]{0mm}{6mm} Contact length, $\ell_c$ & 200 nm & Contact length, $L_c$  & 4 (*)\\
			\hline
			\rule[-2mm]{0mm}{6mm} Mean operating temperature, $T_{op}$ & 532$^\circ$C & Heat loss wire to substrate, $h_w^{(s)}$ & $2.34\times 10^7$ (*,**)\\
			\hline
			    \rule[-2mm]{0mm}{6mm} Surface energy, $\gamma$ & 1500 erg$/$cm$^2$ & Heat loss contact to substrate, $h_c^{(s)}$ & $5.76\times 10^{-4}$ (*,**)\\ 
				\hline
				\rule[-2mm]{0mm}{6mm}  Length of electron mean free path, $s$ & 25 nm & $g=R_0/(2s)$ & 1 (*)\\
				\hline
				\rule[-2mm]{0mm}{6mm}   Wire bulk resistivity, $\rho_b^{(w)}$ & 2.2$\times 10^{-8}$ $\Omega$ cm  & Joule heat, $Q_0$  & $1.83\times 10^{10}$ (*,***)\\
				\hline
				\rule[-2mm]{0mm}{6mm} Ambient temperature, $T_a$ & 20$^\circ$C & Joule heat, $Q_1$ & $4.55\times 10^{-6}$\\ 
			\hline
			\rule[-2mm]{0mm}{6mm}  Applied voltage, $V$  & 0-1 V & Applied voltage, $A$ & 0.1 (see Sec. \ref{Notes}) \\
			\hline
			\rule[-2mm]{0mm}{6mm} Adatom diffusivity activation energy, $\mathcal{E}$ & 1.01 eV & Activation energy, $\mathcal{\bar E}$ & $1.17\times 10^4$ (see Sec. \ref{Notes})\\
			\hline
			\rule[-2mm]{0mm}{6mm} Adatom diffusivity pre-factor, $\mathcal{D}$ & 0.03 cm$^2/$s  & Misorientation angle, $\psi$ & $\pi/12$ \\
			\hline
			\rule[-2mm]{0mm}{6mm}  Adatom surface density, $\nu$ & $5\times 10^{14}$ cm$^{-2}$ & Surface diffusion anisotropy strength, $S$ & 1 \\
			\hline
			\rule[-2mm]{0mm}{6mm} Substrate thermal conductivity, $k_s$ & 0.55 W$/$K m & Thermomigration heat of transport, $H$ & 0-50 (see Sec. \ref{Notes}) \\
			\hline
			\rule[-2mm]{0mm}{6mm} Contact thermal conductivity, $k_c$ & 398 W$/$K m & Contact thermal diffusivity, $D_T^{(c)}$ &  $2.93\times 10^{10}$ (*)\\
			\hline
			\rule[-2mm]{0mm}{6mm} Wire thermal conductivity, $k_w$ & 315 W$/$K m &  &  \\
			\hline
			\rule[-2mm]{0mm}{6mm} Wire specific heat capacity, $c_w$ & 0.13 J$/$g K & &  \\
			\hline
			\rule[-2mm]{0mm}{6mm} Contact specific heat capacity, $c_c$ & 0.39 J$/$g K & &  \\
			\hline
			\rule[-2mm]{0mm}{6mm} Wire mass density, $\chi_w$ & 19.3 g$/$cm$^3$  & &  \\
			\hline
			\rule[-2mm]{0mm}{6mm} Contact mass density, $\chi_c$ & 8.94 g$/$cm$^3$  & &  \\
			\hline
			\rule[-2mm]{0mm}{6mm} Contact thickness, $d_c$ &  50 nm & &  \\ 
			\hline
			\rule[-2mm]{0mm}{6mm} Substrate thickness, $d_s$ &  100-600 nm & &  \\
			\hline
			\rule[-2mm]{0mm}{6mm} Effective charge, $q$ & $10^{-8}$ C & &  \\
			\hline
			\rule[-2mm]{0mm}{6mm} Atomic volume, $\Omega$   & $10^{-22}$ cm$^3$ & & \\
			\hline
\end{tabular}}
\caption[\quad Parameters]{Physical and dimensionless parameters.  Physical parameters are 
for a gold wire, a copper contacts, and a SiO$_2$ substrate. All values are cited at room temperature. 
(*): Value at $R_0=50$ nm and/or $\ell_w=4200$ nm; (**): Value at $d_s=300$ nm; (***): Value at $V=0.2$ V.
}
\label{T1}
\end{table}

\end{document}